\documentclass[fleqn,twoside]{article}
\usepackage{espcrc2}
\usepackage{amsfonts}
\usepackage{amsmath}
\usepackage{amssymb}
\usepackage{graphicx}%
\usepackage[figuresright]{rotating}

\title{Nuclear Lattice Simulations with EFT}
\author{Dean Lee\address{Department of Physics, North Carolina State University, Raleigh, NC, 27603}}

\begin{document}

\begin{abstract}
This proceedings article is a summary of results from work done in
collaboration with Bu\={g}ra Borasoy and Thomas Schaefer. \ We study nuclear
and neutron matter by combining chiral effective field theory with
non-perturbative lattice methods. \ We present results for hot neutron matter
at temperatures 20 to 40 MeV and densities below twice nuclear matter density.

\vspace{1pc}
\end{abstract}
\maketitle

\section{Introduction}

Weinberg extended effective field theory methods to the nucleon-nucleon
interaction \cite{Weinberg:1990rz}, and over the last several years effective
field theory methods have been applied successfully to the two and
three-nucleon system. \ Our aim and
the goal of the Nuclear Lattice Collaboration as a whole \cite{Seki:1998qw}%
\cite{Muller:1999cp} is to extend effective field theory methods to the
nuclear many-body problem. For this purpose we investigate the many-body
physics of low-energy nucleons and pions on the lattice. Our starting point is
the same as that of Weinberg.\ We begin with the most general local Lagrangian
involving pions and low-energy nucleons consistent with translational
invariance, isospin symmetry, and spontaneously broken chiral symmetry. This
yields an infinite set of possible interaction terms with increasing numbers
of derivatives and/or nucleon fields. \ Degrees of freedom associated with
anti-nucleons, heavier mesons such as the $\rho$, and heavier baryons such as
the $\Delta$, are integrated out. \ The contribution of these particles appear
as coefficients of local terms in our pion-nucleon Lagrangian. \ We also
integrate out nucleons with momenta greater than $\pi a^{-1}$, where $a$ is
the lattice spacing.

The operator coefficients in our effective Lagrangian are determined by
fitting to experimentally-measured few-body nucleon scattering data at zero
temperature. \ The dependence on the lattice spacing is described by the
renormalization group and can absorbed by renormalizing operator coefficients.
\ In this way we construct a realistic simulation of many-body nuclear
phenomena with no free parameters. \ In our discussion we present results for
hot neutron matter at temperatures 20 to 40 MeV and densities below twice
nuclear matter density. \ This proceedings article is a summary of results
from work done in collaboration with Bu\={g}ra Borasoy and Thomas Schaefer
\cite{Lee:2004si}.

\section{Non-perturbative effective field theory}

Effective field theory provides a systematic method to compute physical
observables order by order in the small parameter $Q/\Lambda_{\chi}$, where
$\Lambda_{\chi}$ is the chiral symmetry breaking scale and $Q=(q,m_{\pi
},\ldots)$. Here, $q$ is a small external momentum and $m_{\pi}$ is the mass
of the pion. The simplest processes are those that involve only pions and
external fields. In this case the effective field theory is perturbative. At
any order in $Q$ there are only a finite number of diagrams that have to be
included. At lowest order these are tree diagrams with the leading order
interaction. At higher order, diagrams with more loops or higher order terms
in the interaction have to be taken into account.

Weinberg showed \cite{Weinberg:1990rz}\cite{Weinberg:1991um} that the simple
diagrammatic expansion for nucleon-nucleon scattering is spoiled by infrared
divergences. He suggested performing an expansion of the two-particle
irreducible kernel and then iterating the kernel to all orders to produce the
scattering Green's function. It was later pointed out that a possible
difficulty arises because at any order in $Q$ an infinite number of diagrams
is summed, and it is not clear that all the cutoff dependence at that order
can be absorbed into counterterms that are present at that order
\cite{Kaplan:1996xu}. This problem does indeed arise if one considers
nucleon-nucleon scattering in the $^{1}S_{0}$ channel \cite{Beane:2001bc}, but
in practice the cutoff dependence appears to be very weak \cite{Lepage:1997cs}.

In this work we go one step further and consider the nuclear many-body
problem. We expand the terms in our action order by order, $S=S_{0}
+S_{1}+S_{2}+\cdots$. At order $k$ in the chiral expansion, we calculate 
observables by evaluating the functional integral

\begin{equation}
\frac{\int DND\bar{N}D\pi G(\bar{N},N,\pi)e^{-\left[  S_{0}+\cdots
+S_{k}\right]  }}{\int DND\bar{N}D\pi e^{-\left[  S_{0}+\cdots+S_{k}\right]
}}.
\end{equation}

We will refer to this approach as non-perturbative effective field theory.
\ The interactions at chiral order $k$ or less are iterated to arbitrary loop
order. \ The functional integral is computed non-perturbatively by putting the
pion and nucleon fields on the lattice and using Monte Carlo sampling. \ Since
the number of diagrams at a given chiral order grows exponentially with the
number of nucleons, a non-perturbative technique such as this is needed for
systems with more than just a few nucleons.

Computing the path integral corresponds to summing an infinite set of
diagrams. As in the case of iterating the two-particle irreducible kernel to
determine the full two-nucleon Green's function, it is not clear that the
cutoff dependence at a given order in the low energy expansion can be absorbed
into a finite number of coefficients in the action. In practice we will
therefore restrict ourselves to lattice cutoffs that satisfy $\pi
a^{-1}<\Lambda_{\chi}$.

\section{Lowest order interactions}

We let $N$ represent the nucleon fields.  We use $\tau_{i}$ to represent Pauli matrices acting in isospin space, and we
use $\vec{\sigma}$ to represent Pauli matrices acting in spin space. Pion
fields are notated as $\pi_{i}$. \ We denote the pion decay constant as
$F_{\pi}^{phys}\approx183$ MeV and let
$D=1+\pi_{i}^{2}/F_{\pi}^{2}$.  The lowest order Lagrange density for low-energy pions and nucleons is given
by \cite{Ordonez:1996rz},

\begin{align}
\mathcal{L}^{(0)} &  =-\tfrac{1}{2}D^{-2}\left[  (\vec{\nabla}\pi_{i}%
)^{2}-\dot{\pi}_{i}^{2}\right]  -\tfrac{1}{2}D^{-1}m_{\pi}^{2}\pi_{i}%
^{2}\nonumber\\
&  +\bar{N}[i\partial_{0}-(m_{N}-\mu)]N\nonumber\\
&  -D^{-1}F_{\pi}^{-1}g_{A}\bar{N}\left[  \tau_{i}\vec{\sigma}\cdot\vec
{\nabla}\pi_{i}\right]  N\nonumber\\
&  -D^{-1}F_{\pi}^{-2}\bar{N}[\epsilon_{ijk}\tau_{i}\pi_{j}\dot{\pi}%
_{k}]N\nonumber\\
&  -\tfrac{C_{S}}{2}\bar{N}N\bar{N}N-\tfrac{C_{T}}{2}\bar{N}\vec{\sigma
}N\cdot\bar{N}\vec{\sigma}N.
\end{align}

$g_{A}$ is the nucleon axial coupling, and $\epsilon_{ijk}$ is the Levi-Civita
symbol. \ The chemical potential $\mu$ controls the nucleon density and $\mu$
will be set very close to $m_{N}$. \ At next order we have terms%
\begin{equation}
\mathcal{L}^{(1)}=\tfrac{1}{2m_{N}}\bar{N}\vec{\nabla}^{2}N+...
\end{equation}
We\ will include this kinetic energy term from $\mathcal{L}^{(1)}$ in our
lowest-order Lagrange density so that we get the usual free nucleon propagator.  In this study we limit ourselves to the interactions of neutrons and neutral
pions.

\section{Lattice Schr\"{o}dinger equation and phase shifts}

We adjust $C$, the coefficient of the $\bar{N}N\bar{N}N$ contact interaction,
so that the $NN$ s-wave scattering length matches the experimental value (see
for example \cite{Funk:2000mf}). \ In order to calculate the phase shifts, we
will solve the lattice Schr\"{o}dinger equation for the two-neutron system and
observe the asymptotic form of the scattering wavefunctions.

With the lowest order two-nucleon potential we can construct a matrix
representation for the Hamiltonian in the two-neutron sector and solve the
time-independent lattice Schr\"{o}dinger equation. \ At this stage one could
implement L\"{u}scher's formula for measuring phase shifts in a cubical
periodic box \cite{Luscher:1991ux}\cite{Luscher:1986pf}. \ However in our case
we can construct the eigenvectors explicitly using Lanczos iteration, and so
we find it more straightforward and accurate to read the phase shifts directly
from the asymptotic forms of the s-wave scattering states.  We use the lattice Schr\"{o}dinger technique to tune the coupling $C$ to
reproduce the large scattering length that is observed in nature. 

\section{Results}

We have generated simulation results for $a^{-1}=150$ MeV; $\alpha_{t}%
=\tfrac{a_{t}}{a}=1.0$; temperatures $T^{phys}=25.0$ MeV and $37.5$\ MeV; and
lattice sizes $3^{3}$, $4^{3}$, and $5^{3}$. \ Half-filling at this lattice
spacing occurs at $\rho=2.64\rho_{N}^{phys}$, where $\rho_{N}^{phys}$ is the normal nuclear density of about $0.17$ nucleons
per fm$^{3}$. \ The calculations were performed using the zone determinant
method \cite{Lee:2003mb} using the second order approximation with zones
consisting of a single spatial point. \ By calculating the exact determinants
of some generated matrix configurations we estimate the systemic error for the
zone expansion to be about $<0.1\%$ for $T^{phys}=37.5$ MeV and $<0.5\%$ for
$T^{phys}=25.0$ MeV.

We have dealt with the complex action by computing the phase as an observable,
and for the various simulations presented here we found an average phase of
about $\sim0.95$ and so this did not present a significant computational problem.%

\begin{figure}
[ptb]
\begin{center}
\includegraphics[
width=2.00in,angle=-90
]%
{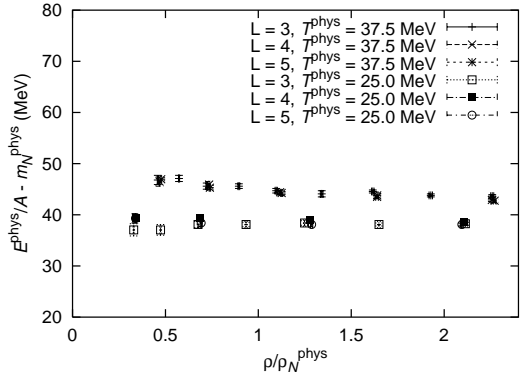}%
\caption{Energy per neutron in MeV for temperatures 25 MeV and 37.5 MeV and
different lattice volumes. \ The inverse lattice spacing is $a^{-1}=150$ MeV
and $\alpha_{t}=1.0$.}%
\label{ldependence}%
\end{center}
\end{figure}
In Fig. \ref{ldependence} we show the energy per neutron as a function of
neutron density. \ Our results indicate a rather flat function for the energy
per neutron as a function of density.

The flatness of our energy per neutron curves at these temperatures are
intriguing and hopefully will be checked by others in the near future. \ The
results of \cite{Muller:1999cp} also see a flattening of the energy per
neutron curve with increasing temperatures. \ We can also compare with
variational calculations \cite{Friedman:1981qw} and recent quantum Monte Carlo
results from \cite{Carlson:2003wm}. They observe a significant flattening of
the energy per neutron curve due to iteractions even at zero temperature.

\end{document}